**GMN4AD: Graph Matching Network for Alzheimer's Disease Diagnosis with Test-Time Domain Adaptation using Multi-centered Structure Magnetic Resonance Imaging**

Chen Zhao[1], Huan Huang[1], Yixin Xie[2], Jiajing Huang[3], Weihua Zhou[4]

1. Department of Computer Science, Kennesaw State University, Marietta, GA, 30060

2. Department of Information Technology, Kennesaw State University, Marietta, GA, 30060

3. School of Data Science and Analytics, Kennesaw State University, Marietta, GA, 30060

4. Department of Applied Computing, Michigan Technological University, Houghton, MI, 49931, USA

Correspondence: Chen Zhao, czhao4@kennesaw.edu

**Abstract**

Alzheimer's Disease (AD) is a progressive neurodegenerative disorder that affects millions of older adults, with prevalence expected to rise significantly in the coming years. Early diagnosis, particularly during the mild cognitive impairment (MCI) stage, is critical for timely intervention. Structural Magnetic Resonance Imaging (sMRI) has emerged as a key modality for detecting AD-related brain changes, but traditional graph-based approaches often struggle with modality and inter-site heterogeneity, limiting diagnostic performance. In this paper, we propose Graph Matching Network for Alzheimer's Disease Diagnosis (GMN4AD), designed to model interactions between heterogeneous brain graphs derived from neuroimaging data. Unlike conventional methods that treat each brain graph independently, GMN4AD leverages graph matching to capture cross-graph relationships, enhancing diagnostic precision. Furthermore, we introduce a test-time domain adaptation strategy that combines contrastive learning to mitigate domain shifts during inference. Extensive experiments on three public AD datasets demonstrate that GMN4AD achieves superior performance compared to state-of-the-art methods, offering a robust and generalizable solution for AD diagnosis.

## 1. Introduction

Alzheimer's Disease (AD) is a neurodegenerative disorder associated with aging that leads to progressive memory loss and cognitive decline [1]. According to the CDC report in 2022, 11.3% of older adults in the United States reported experiencing subjective cognitive decline or worsening memory loss. Specifically, 10.3% of individuals aged 50 to 64 and 12.3% of those aged 65 and older reported these symptoms, with notable ethnic disparities in risk factors among those affected by AD [2]. In addition, according to estimates from the Alzheimer's Association, nearly 8 million people in the United States aged 65 years and older are expected to develop Alzheimer's disease by 2030 [3]. While aging is recognized as a key risk factor, the underlying biological mechanisms of AD remain incompletely understood, and no current treatments effectively halt the disease's progression [4].

Identifying AD along a continuum that includes cognitive normal (CN), mild cognitive impairment (MCI), and full-blown AD allows for earlier interventions [5]. MCI, which can be a precursor to AD with a conversion rate of 12%-15% per year [6], is a critical phase where cognitive decline may be detected before significant functional impairment occurs [7]. Patients in the MCI stage might benefit from monitoring and supportive therapies, while those diagnosed with AD may require more intensive management strategies.

Magnetic Resonance Imaging (MRI) is the key imaging modalities in AD risk prediction. Structural MRI is a valuable tool for identifying anatomical changes associated with AD, such as brain atrophy and other pathological features, thereby aiding in clinical diagnosis, achieving a sensitivity range from 65% to 80% and a specificity between 76% to 85% [8–10]. In addition, the diagnosis of AD benefits significantly from the use of connectivity between biomarkers [11]. Under this scenario, a graph model can effectively capture and represent these complex interactions, where nodes in the graph can represent different biomarkers, such as MRI atrophy regions [12]. And edges illustrate relationships and correlations between these biomarkers, such as how changes in MRI atrophy correlate. In this paper, we aim to build graph-based models that enhance AD diagnosis and reveal data interactions to identify key biomarkers.

Modality heterogeneity results in difficulties in accurately capturing the complex relationships and limited performance for AD diagnosis [13]. However, differences in scanner settings, acquisition protocols, and population characteristics introduce inter-site heterogeneity, thereby degrading cross-site generalization. A significant challenge in AD diagnosis using traditional multi-modality-based methods lies in the heterogeneity of the graph structures derived from MRI data [12,14–16]. These methods typically treat imaging modalities as individual graphs and apply graph neural networks (GNNs) to perform diagnosis. However, due to differences in imaging resolutions, patient variability, and the complex nature of brain connectivity, the resulting graphs often exhibit structural and feature-level heterogeneity. This mismatch can lead to suboptimal performance, as GNNs struggle to fully capture the intricate relationships and interactions between the graphs [17].

In this research, we propose **G**raph **M**atching **N**etwork for **A**lzheimer's **D**isease Diagnosis (**GMN4AD**), a heterogeneous graph matching–based framework designed to model interactions between distinct brain graphs rather than treating them as isolated entities, as shown in Figure 1. This design enables a more precise characterization of the brain network derived from neuroimaging data associated with different status of AD, thereby improving diagnostic accuracy. GMN4AD is evaluated on three public datasets, achieving superior performance across various metrics.

The main contributions of this work are summarized as follows:

1) Graph matching-based diagnosis: Instead of modeling each brain as an independent graph, we introduce a graph matching–based method that compares features across different brain graphs for AD diagnosis.

2) Test-time domain adaptation: We propose a test-time domain adaptation strategy that integrates contrastive learning and graph reconstruction loss to address domain shifts during inference, thereby mitigating performance degradation.

3) Comprehensive evaluation: Extensive experiments on three public AD-related datasets demonstrate that GMN4AD achieves competitive or superior performance compared to state-of-the-art methods.

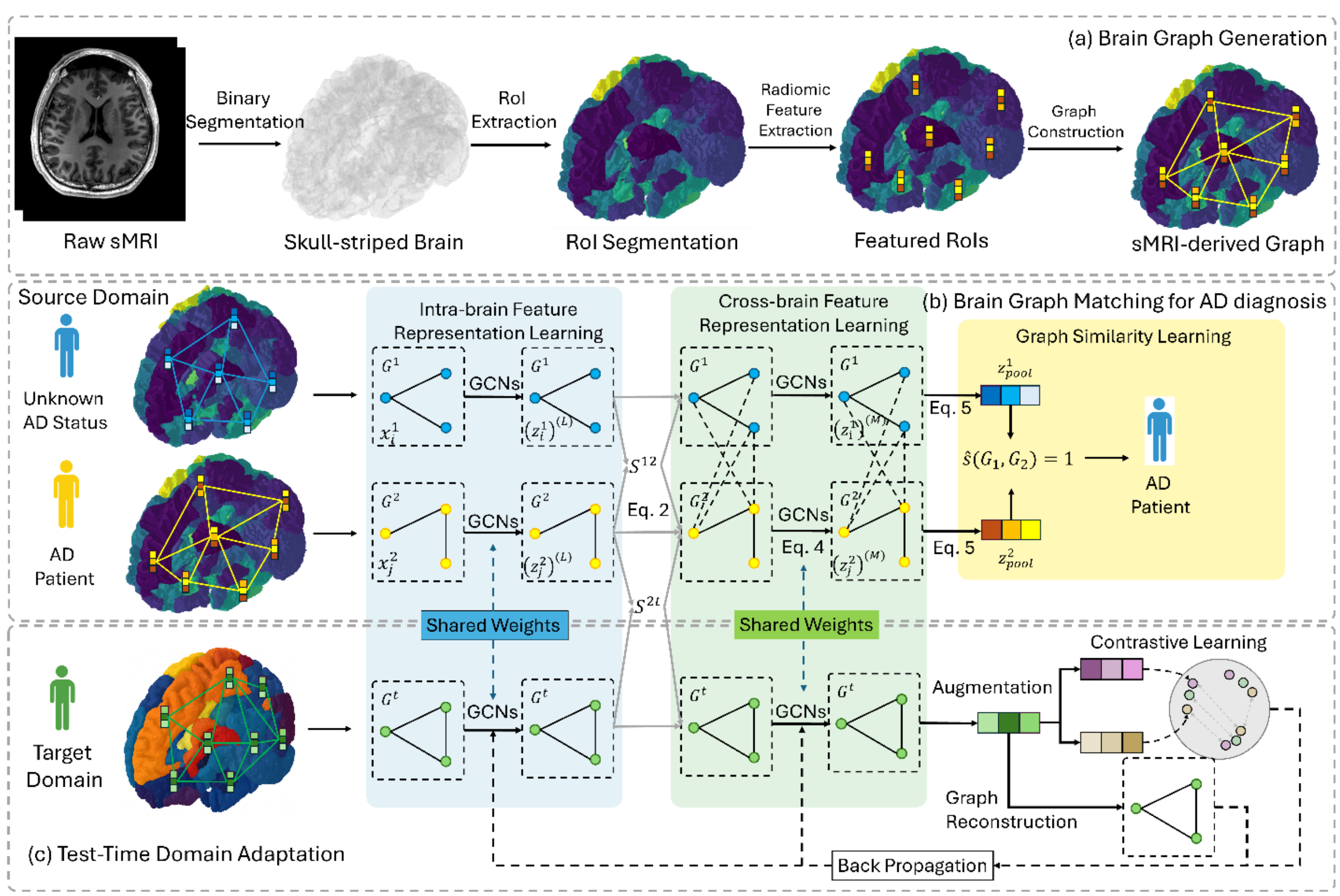

**Figure 1.** Overview of the proposed graph matching–based neural network for AD diagnosis with test-time domain adaptation. (a) Patient-specific brain graph generation using structural MRI (sMRI): Each brain graph is constructed from sMRI data, where colored rectangles represent radiomic features extracted from the corresponding regions of interest (ROIs), and yellow edges indicate brain connectivity; (b) Graph matching for AD diagnosis in the source domain: Two example patients are shown. The blue node denotes an unseen (test) patient, while the yellow node represents a patient with a known AD status. By performing graph matching between the two, the proposed GMN4AD quantifies graph similarity and infers the AD status of the test patient based on the matched reference. (c) Test-time domain adaptation on the target dataset: Instead of directly applying the source-trained model to the target domain, an unsupervised test-time domain adaptation strategy is employed. The model is further fine-tuned using a combination of contrastive learning loss and graph reconstruction loss. Without requiring target-domain labels, the weights of GMN4AD are adaptively updated to improve generalization and robustness under domain shifts.

## 2. Related Work

### 2.1. AD diagnosis using graph neural networks

GNNs have emerged as a powerful framework for AD diagnosis by effectively modeling the complex relationships among brain regions [18]. In AD diagnosis field, it categorizes subject-level graph and population-level graph. In population-level graph approaches, each node represents a patient, and edges capture similarity or connectivity between patients based on imaging or phenotypic features. This setup allows GNNs to model inter-subject relationships and capture population-level patterns relevant to disease progression. Recent studies have explored various strategies for learning from population graphs, including dual-attention GNNs that enhance representation learning through self- and cross-attention [19], multi-branch GNNs that process patient graphs and combine embeddings at multiple levels [20], and spectral graph attention with adaptive fusion to model intra- and inter-cluster relationships [21]. Collectively, these methods demonstrate that population-graph GNNs provide a powerful and interpretable framework for AD diagnosis [22]. Despite their effectiveness, population-level graph GNNs for AD diagnosis often suffer from sensitivity to graph construction, scalability challenges, limited interpretability, and vulnerability to patient heterogeneity or missing data.

Unlike conventional machine learning methods that treat neuroimaging features as independent variables, subject-level GNNs represent the brain as a graph, where nodes correspond to regions of interest (ROIs) and edges capture structural or functional connectivity [23]. This graph-based representation enables GNNs to integrate both local features and global topological information, allowing for more biologically meaningful learning [24]. By aggregating information across connected regions, GNNs can identify subtle and spatially distributed patterns associated with AD progression, thereby improving the accuracy and interpretability of disease classification. Recent studies have demonstrated that GNN-based approaches outperform traditional deep learning models in distinguishing between CN, MCI, and AD subjects, highlighting their potential for early diagnosis and clinical decision support [25].

### 2.2. Graph matching and graph similarity learning

Graph similarity learning has emerged as a powerful paradigm for modeling relationships between complex graph-structured data, with growing relevance in neuroimaging analysis for AD diagnosis. Conventional similarity estimation methods rely on handcrafted metrics that often fail to capture the intricate structural and functional dependencies within brain networks [26]. In contrast, GNNs have become a cornerstone of these methods due to their ability to integrate both local and global graph information [27]. Siamese GNNs, for instance, learn embeddings for paired brain graphs to quantify structural similarity between subjects [28], while graph matching networks employ attention-based mechanisms to align nodes across brain regions, enabling more precise correspondence modeling [29]. Additionally, deep graph kernel methods map graphs into a latent feature space, allowing similarity to be measured through learned distance metrics that combine the interpretability of kernels with the flexibility of deep representations [27]. However, existing methods often fail to capture the subtle and heterogeneous alterations in brain connectivity associated with AD. Our GMN4AD integrates domain-informed graph representations with cross-graph matching to enhance both diagnostic accuracy and interpretability in characterizing subject-level brain network similarities.

### 2.3. Domain adaptation and test-time adaptation

Test-time adaptation (TTA) has recently gained attention as a technique that allows pre-trained models to adjust to unseen, unlabeled test data before making predictions [30]. Several strategies have been developed to implement TTA, including self-supervised learning [31,32], updating batch normalization statistics [33,34], and adapting input data [35–37] to improve model robustness and inference accuracy. Despite these advances, TTA has not yet been explored in graph matching and graph similarity learning, and its potential for AD diagnosis using multi-centered datasets remains largely unexplored.

## 3. Methodology

### 3.1 Modeling sMRI in Graph

To model the structural relationships between brain regions for AD diagnosis, we construct brain graphs from MRI scans, as shown in Figure 1 (a). In the subject-level graph, each node represents a brain ROI segmented by FreeSurfer [38], and edges encode spatial or functional relationships between regions, capturing the anatomical connectivity of the brain. For each ROI, voxel coordinates are used to compute the centroid, which is converted to real-world coordinates via the MRI affine transformation. Radiomics features [39,40] , including texture, intensity, and shape, are extracted from the cropped ROI volume to form node-level feature vectors. These features are denoted as $x_i^g \in \mathbb{R}^d$, where $i \in \{1,2,\cdots,n\}$ represents the node index and $d$ is the feature dimension. Specifically, $d = 113$ features are extracted, as detailed in Table 1.

The brain graph is defined as $G = (V, E)$, where $V = \{V_1, V_2, \cdots V_n\}$ denotes the set of ROIs, $E = \{E_1, E_2, \cdots, E_{n_e}\}$ represents edges, which are derived from the spatial relationships between ROI centroids. $n$ represents the number of nodes, as well as the number of ROIs in the sMRI, and $n_e$ is the number of edges in the MRI derived graph. The adjacency matrix form of $G$ is $A = \left(a_{ij}\right)_{i,j=1}^{n} \in \{0,1\}^{n\times n}$, where.

$$a_{ij} = \begin{cases} 1, & if\ there\ is\ an\ edge\ from\ node\ i\ to\ node\ j \\ 0, & otherwise \end{cases} \quad (1)$$

**Table 1.** Feature for each node (ROI) in the sMRI-derived graph

| Index | Feature explanation |
|---|---|
| 1-5 | Mean, minimum, and maximum intensity values of the ROI, along with the voxel count and volume. |
| 6-16 | Geometric properties of the ROI, including elongation, axis lengths, maximum 2D diameters, mesh area, minor axis length, sphericity, surface area, surface-to-area ratio, and pixel-based area. |
| 17-34 | First-order radiomics features, which capture the intensity distribution within the ROI, including percentile values, energy, entropy, statistical dispersion, and measures of central tendency (mean, median), providing quantitative insights into brightness variations. |
| 35-58 | Gray Level Co-occurrence Matrix (GLCM) features, which quantify textural patterns within the ROI by measuring spatial relationships between pixel intensities. |
| 59-73 | Gray Level Dependence Matrix (GLDM) features, which characterize texture by measuring the distribution and variance of dependent gray levels. |
| 74-89 | Gray Level Run Length Matrix (GLRLM) features quantify texture patterns by analyzing the distribution of consecutive pixels with the same intensity. |
| 90-105 | Gray Level Size Zone Matrix (GLSZM) features, which evaluate textural patterns based on the size and distribution of homogeneous zones with the same gray level. |
| 106-110 | Neighborhood Gray Tone Difference Matrix (NGTDM) features, which assess texture by analyzing the variation in gray-tone differences between neighboring pixels. |
| 111-113 | x,y and z, indicating the relative coordinates of the center mass for each brain ROI. |

### 3.2. Iterative Cross Graph Matching for AD detection

The sMRI-derived structured brain graph serves as input for graph matching, where graphs of unknown-status subjects are compared against those of confirmed AD patients. By analyzing cross-graph similarities, Alzheimer's disease progression can be effectively assessed, providing a novel and interpretable framework for AD diagnosis. Formally, given two sMRI-derived graphs, $G^1 = (V^1, E^1)$ and $G^2 = (V^2, E^2)$, the proposed GMN4AD aims to calculate the structural similarity between $G^1$ and $G^2$. If the sMRI corresponding to $G^1$ comes from a confirmed AD patient, and $G^2$ comes from a subject with unknown AD status, then the similarity between $G^1$ and $G^2$ reflects the degree to which the unknown subject's brain structure resembles that of the AD patient. A higher similarity indicates stronger AD-like structural degeneration; conversely, a lower similarity suggests preserved connectivity and a healthier brain pattern. The overall graph-matching network contains the following five modules, as shown in Figure 1.

1) *Intra-graph Feature Embedding*. The intra-graph features embedding module employs GCNs and a multi-layer perceptron (MLP) to learn expressive representations of each brain region node within the sMRI-derived brain graph $G$. This module aggregates both local topology and feature information to capture intra-regional dependencies and structural variations relevant to Alzheimer's pathology. Formally, for a node $V_i$ in graph $G$, the intra-graph feature embedding is represented as Eq. 1.

$$z_i^{(l+1)} = f_{intra_emb}^{(l)}\left(\left[\sum_{j\in E_i} z_j^{(l)}, z_i^{(l)}\right]\right) \tag{2}$$

where $l \in \{1,\cdots,L\}$ denotes the GCN layer index; $f_{intra_emb}^{(l)}$ is the MLP for the $l$-th layer; $E_i$ represents the set of edges connected to node $V_i$, and $\sum_{j\in E_i}\cdot$ indicates the element-wise aggregation of features from neighboring nodes, as determined by the adjacency matrix of $G$. When $l = 1$, the node feature is initialized as $z_i^1 = x_i \in \mathbb{R}^d$. The square brackets $[\cdot]$ denote feature concatenation. Eq. 1 thus models the message-passing and neighborhood aggregation process within the GCN to embed intra-graph structure information. The resulting representation $z_i^{(L)}$ captures both the local connectivity and the intrinsic radiomic characteristics of brain region $V_i$.

To establish structural correspondence between two brain graphs $G^1$ and $G^2$, node-to-node affinities are computed in the embedding space using their updated representations. The intra-graph node similarity matrix is obtained via a weighted dot product, followed by normalization through the Sinkhorn operator [41], as shown in Eq. 2.

$$S_{ij} = Sinkhorn\left(\exp\left(\frac{(z_i^1)^{(L)} \cdot A_{intra} \cdot \left(z_j^2\right)^{(L)}}{\sqrt{d_{intra}}}\right)\right) \tag{3}$$

where $i$ and $j$ are the node indices in $G^1$ and $G^2$; $d_{intra}$ is the feature dimension, and $A_{intra} \in \mathbb{R}^{d_{intra}\times d_{intra}}$ is a learnable parameter matrix that encodes node affinities. The exponential operation ensures non-negative affinity scores required by the Sinkhorn algorithm [41]. Eq. 2 integrates GCN-based embeddings with the Sinkhorn normalization to achieve consistent and interpretable graph alignment across subjects [42].

2) *Cross-graph Feature Embedding in Graph Pairs*. Cross-graph feature embedding plays a crucial role in improving the reliability of node correspondence between sMRI-derived brain graphs. Without interactive information exchange between graphs, direct node-to-node matching can be unstable and sensitive to noise [43]. This module enables information propagation across two distinct graphs, allowing the model to learn richer inter-graph dependencies and enhance structural alignment accuracy [44]. Formally, for two brain graphs $G^1$ and $G^2$, the cross-graph feature embedding for node $V_i$ in $G^1$, considering all nodes in both graphs, is defined as Eq. 3.

$$(z_i^1)^{(m+1)} = f_{cross}^{(m)}\left(\left[\sum_{j=1}^{n_2} S_{ij} \cdot \left(z_j^2\right)^{(m)}, (z_i^1)^{(m)}\right]\right) \tag{4}$$

where $f_{cross_emb}^{(m)}$ denotes the MLP used in the $m$-th layer of the cross-graph embedding module, and $m \in \{1,\cdots,M\}$. When $m = 1$, $(z_i^1)^{(1)}$ corresponds to $(z_i^1)^{(L)}$, the final intra-graph feature embedding obtained from the previous module. Cross-graph embedding is iteratively applied $M$ times to extract hierarchical and non-linear features that capture structural correspondences between brain regions across subjects. Symmetrically, for node $V_j$ in $G^2$, the updated cross-graph feature embedding is calculated as shown in Eq. 4.

$$\left(z_j^2\right)^{(m+1)} = f_{cross}^{(m)}\left(\left[\sum_{i=1}^{n_1} S_{ij}\cdot\left(z_i^1\right)^{(m)}, \left(z_j^2\right)^{(m)}\right]\right) \tag{5}$$

After applying both intra-graph and cross-graph feature embedding, each node $V_i$ obtains a refined representation $\left(z_i^g\right)^{(M)} \in \mathbb{R}^{d_{cross}}$, where $g \in \{1,2\}$, and $d_{cross}$ denotes the dimensionality of the learned cross-graph feature space.

3) *Global feature extraction*. Following the multi-level feature embedding process, a graph pooling operation is applied to summarize each brain graph into a compact global representation. Formally, for graph $G^g$, the pooled feature vector [45] is defined as Eq. 5.

$$z_p^g = \frac{1}{n}\sum_{i=1}^{n}\left(z_i^g\right)^{(M)} \tag{6}$$

where $z_p^g \in \mathbb{R}^{d_{cross}}$ represents the average feature vector describing the overall structure of brain graph $G^g$. The pooling operation summarizes the information from all brain regions by averaging their node embeddings, thereby transforming variable-sized brain graphs into compact, fixed-dimensional representations suitable for subsequent similarity computation and disease classification.

4) *Graph similarity measurement*. After graph pooling, the similarity between the two sMRI-derived brain graphs is evaluated to quantify their global structural correspondence. The cosine similarity between the pooled feature embeddings of $G^1$ and $G^2$ is adopted as the graph-level similarity metric, which effectively captures overall topology and feature consistency. This measure reflects how closely the subject with unknown AD status resembles the confirmed AD patient in the learned embedding space. Formally, it is defined as Eq. 6. This approach is both efficient and effective when the node features carry significant information.

$$\hat{s}(G_1, G_2) = \frac{z_p^1 \cdot z_p^2}{\left\|z_p^1\right\|\left\|z_p^2\right\|} \tag{7}$$

where $z_p^1$ and $z_p^2$ denote the pooled feature vectors of the two graphs $G^1$ and $G^2$. The resulting similarity score $\hat{s} \in [0,1]$ represents the degree of structural alignment between $G^1$ and $G^2$. If they are from the same AD status, then the ground truth of $\hat{s}$ is 1; otherwise, the ground truth of $\hat{s}$ is 0.

### 3.3. Test-time Domain Adaptation

To improve model generalization under domain shifts, we propose a test-time domain adaptation (TTA) strategy for the cross-graph graph matching model. Unlike standard transfer learning, TTA adapts the model online during inference without requiring labeled target-domain data. Specifically, given a target patient graph, we fine-tune the model using a contrastive consistency loss computed on both the source and target domains.

To enhance robustness and promote stable representations, we introduce random feature perturbations to generate augmented node embeddings. Given the node embedding matrix after cross-graph feature embedding, the augmented embedding $z_i'$ is obtained by adding small Gaussian noise and applying random edge dropout, as shown in Eq. 7.

$$z_i' = \tilde{A}_{intra} z_i + \epsilon, \epsilon \in \mathcal{N}(0, \sigma^2 I) \tag{8}$$

where $\tilde{A}_{intra}$ denotes the adjacency matrix after random edge dropout, and $\sigma$ controls the noise level. This perturbation encourages the model to learn invariant representations under minor structural and feature-level variations. A graph-level representation is obtained by averaging node embeddings, and a contrastive loss is applied to align the original and augmented embeddings in Eq. 8.

$$\mathcal{L}_{\text{con}} = -\frac{1}{N}\sum_{i} \log \frac{\exp\left(\text{sim}(z_i, z_i')/\tau\right)}{\sum_j \exp\left(\text{sim}(z_i, z_j')/\tau\right)} \quad (9)$$

where $z_i$and $z_i'$ are the graph-level embeddings of the original and augmented graphs, $\text{sim}(\cdot,\cdot)$ is the cosine similarity, and $\tau$ is a temperature hyperparameter.

The total TTA loss is a weighted sum of these two terms, as shown in Eq. 9.

$$\mathcal{L}_{\text{TTA}} = \mathcal{L}_{\text{con}}^{\text{src}} + \mathcal{L}_{\text{con}}^{\text{tgt}} \quad (10)$$

where $\mathcal{L}_{\text{con}}^{\text{src}}$ and $\mathcal{L}_{\text{con}}^{\text{tgt}}$ indicate the contrastive loss for the graph pairs used in the source domain and target domain, respectively. During the inference, the GMN4AD model is fine-tuned online on each target patient graph using $\mathcal{L}_{\text{TTA}}$, producing graph embeddings better aligned with the target domain and improving AD diagnostic performance.

### 3.4. Model Training and Testing

**Model training:** During model training, two separate graphs with an equal number of ROIs in the sMRI were randomly chosen to form a graph matching pair. The GM4AD is trained using binary cross entropy as the loss function, as shown in Eq. 10.

$$L_{BCE} = -\frac{1}{B}\sum_{b=1}^{B}\left(s(G_1^b, G_2^b) \cdot \log \hat{s}(G_1^b, G_2^b) + (1 - s(G_1^b, G_2^b)) \cdot \log(1 - \hat{s}(G_1^b, G_2^b))\right)^2 \quad (11)$$

where $B$ indicates batch size, $\hat{s}(G_1^b, G_2^b)$ is the measured brain graph similarity, and $s(G_1^b, G_2^b)$ indicates the ground truth. If $G_1^b$ and $G_2^b$ come from same AD status, then $s(G_1^b, G_2^b) = 1$; otherwise $s(G_1^b, G_2^b) = 0$.

**Model testing**: During testing, each graph in the test set is compared against all graphs in a template set to measure their pairwise similarity using Eq. 6. If two graphs originate from subjects with the same AD status, the computed similarity $\hat{s}(G_{test}, G_{template})$ is expected to be close to 1; otherwise, it should approach 0. For each test graph, a set of similarity scores is obtained by matching it with every template graph. A majority voting strategy is then employed to determine the final prediction: template samples with similarity scores above a predefined threshold (e.g., 0.95) are considered matches, and the predicted label of the test graph is assigned to the most frequent AD status among these matched templates. This voting-based inference allows robust diagnosis by aggregating multiple similarity comparisons rather than relying on a single pairwise decision. The pseudo code of training and testing is shown in Algorithms 1 and 2.

**Algorithm 1**. Training algorithm for GMN4AD.

**Input:**

- Training graphs: $\mathcal{G}_{train} = \{G_1^{train}, G_2^{train}, \cdots, G_N^{train}\}$
- Batch size: B, learning rate: $\eta$

**Output:** Trained GMN4AD with model parameters $\theta$

1: Initialize model parameters $\theta$

2: **while** not converged do

3: Randomly sample $B$ pairs $(G_i^b, G_j^b)$ from $\mathcal{G}_{train}$

4: **for** each pair $(G_i^b, G_j^b)$ do

5: Compute predicted similarity: $\hat{s}(G_1^b, G_2^b) = GM4AD(G_1^b, G_2^b \mid \theta)$

Define ground truth label $s(G_1^b, G_2^b) = 1$ if both graphs share same AD status; 0 otherwise.

6: Compute binary cross-entropy loss defined in Eq. 10

7: **end for**

8: Update model parameters by $\theta = \theta - \eta \nabla_\theta L_{BCE}$

9: **end while**

---

**Algorithm 2**. Testing algorithm for GMN4AD.

**Input:**

- Template graphs: $\mathcal{G}_{template} = \{G_1^{template}, G_2^{template}, \cdots, G_T^{template}\}$
- Test graph: $G_{test}$
- Similarity threshold: $t$

**Output:** Predicted AD status for $G_{test}$

Testing:

1: Initialize empty list $S_test$

2: **for** each template graph $G_i^{template} \in \mathcal{G}_{template}$ do

3: Compute predicted similarity: $\hat{s}(G_{test}, G_i^{template}) = GM4AD(G_{test}, G_i^{template} \mid \theta)$

Identify matched templates: $M = \{ G_i^{template} \mid \hat{s}(G_{test}, G_i^{template}) \geq t\}$

4: **end for**

5: Determine predicted label: $label(G_{test}) = \text{mode}(\text{ AD_status}(G_i^{template}) \text{ for } G_i^{template} \in M)$

---

**Test time domain adaptation**: In the test-time domain adaptation phase, we fine-tune the pre-trained GMN4AD model using unlabeled target-domain data to mitigate distributional discrepancies between the source and target datasets. This process follows an unsupervised domain adaptation paradigm, where no target labels are used during adaptation. Specifically, only the parameters within the graph neural and cross-graph attention layers are updated, while all other components remain frozen. The model is optimized using the Adam optimizer with a learning rate of $\eta_{\text{TTA}}$, and the adaptation loss $L_{\text{TTA}}$ defined in Eq. 9 is computed during training. The target dataset is divided into training and testing subsets according to a predefined ratio, and mini-batches from both the source and target training subsets are jointly used for iterative adaptation. After each adaptation epoch, the model's performance is evaluated on the target testing set by measuring the similarity between graphs in the target testing set and template graphs from the source dataset, followed by majority-voting–based classification as in the standard inference stage. The model achieving the highest F1 score is retained as the final adapted version, with early stopping applied to prevent overfitting.

The detailed TTA training procedure is shown in Algorithm 3. Once the model parameters are adapted to target dataset, the majority voting strategy similar to Algorithm 2 is applied to the testing samples in the target dataset using the template graphs in the *source* dataset.

---

**Algorithm 3**. Test-Time Domain Adaptation of GMN4AD.

**Input:**

- Pre-trained GMN4AD model with parameters $\theta$

- Training graphs in source dataset: $\mathcal{G}_{train} = \{G_1^{src}, G_2^{src}, \cdots, G_N^{src}\}$
- Template graphs in source dataset: $\mathcal{G}_{template} = \{G_1^{template}, G_2^{template}, \cdots, G_T^{template}\}$
- Training graphs in target dataset: $\mathcal{G}_{tgt} = \{G_1^{tgt}, G_2^{tgt}, \cdots, G_M^{tgt}\}$
- Similarity threshold: $t$; learning rate for TTA: $\eta_{TTA}$;

**Output:** TTA adapted GMN4AD with model parameters $\theta'$

1: **while** not converged do

2: Randomly sample $B$ pairs $\left(G_i^{src}, G_j^{tgt}\right)$, where $i \in [1, \cdots, N]$ and $j \in [1, T]$

3: **for** each pair $\left(G_i^{src}, G_j^{tgt}\right)$ do

4: Compute predicted feature embedding $z_{pool}^{src}$ and $z_{pool}^{tgt}$ using Eq. 6.

5: Compute contrastive loss defined in Eq. 9.

6: **end for**

7: Update model parameters by $\theta' = \theta - \eta_{TTA} \nabla_\theta L_{TTA}$

8: **end while**

**Evaluation**: To comprehensively assess the diagnostic performance of our framework, classification metrics, accuracy (ACC), precision (PREC), sensitivity (SEN), F1-score (F1), and specificity (SPEC), are measured, and corresponding equations are shown below:

$$\mathrm{ACC} = \frac{\mathrm{TP} + \mathrm{TN}}{\mathrm{TP} + \mathrm{TN} + \mathrm{FP} + \mathrm{FN}} \tag{12}$$

$$\mathrm{PREC} = \frac{\mathrm{TP}}{\mathrm{TP} + \mathrm{FP}} \tag{13}$$

$$\mathrm{SEN} = \frac{\mathrm{TP}}{\mathrm{TP} + \mathrm{FN}} \tag{14}$$

$$\mathrm{F1} = \frac{2 \times \mathrm{PPV} \times \mathrm{TPR}}{\mathrm{PPV} + \mathrm{TPR}} \tag{15}$$

$$\mathrm{SPEC} = \frac{\mathrm{TN}}{\mathrm{TN} + \mathrm{FP}} \tag{16}$$

Here, TP, TN, FP and FN are true positive, true negative, false positive and false negative. Accuracy measures the overall correction of positive and negative predictions; precision (positive predictive value) measures the portion of true positive among predicted positive; sensitivity (recall, or true positive rate) measures the portion of positive correctly predicted among true positive; F1-score is the harmonic means of precision and sensitivity, providing a balanced assessment of performance under potential class imbalance; specificity (true negative rate) measures the portion of negative correctly predicted among true negative. In our study, we conduct three different classification tasks, namely AD vs. CN, MCI vs. CN, and AD vs. MCI. For the first two tasks (AD vs. CN, MCI vs. CN), CN is negative while diseased group (AD/MCI) is positive; For the last task (AD vs. MCI), AD is positive and MCI is negative. Together, these metrics offer a comprehensive evaluation of the model's effectiveness in distinguishing AD from non-AD subjects across different testing scenarios.

### 3.5. Interpretability of GMN4AD

We employ a perturbation-based explanation method, ZORRO [46], to interpret both node and feature importance in our GMN4AD model. Standard ZORRO uses discrete node and feature masks to identify the elements most influential for a model's predictions: for a given input graph, it iteratively and recursively selects important nodes and features based on a fidelity score, which measures the change in the model's prediction after masking them [47]. As a post-hoc method, ZORRO does not require retraining the model or a differentiable (continuous) mask; instead, it uses a hard mask, where 1 indicates selection and 0 indicates exclusion.

While the original ZORRO explains individual graphs independently, our task involves paired graph matching, where the model predicts similarity between two graphs. To address this, we extend ZORRO to a paired explanation setting, producing shared node and feature masks across both graphs in the pair. This modification ensures interpretability in the context of graph similarity, highlighting the nodes and features in each graph that jointly drive the model's prediction. By applying this *ZORRO-Paired* strategy, we can visualize and understand which brain regions and corresponding radiomic features are most critical for determining similarity in AD-related brain networks.

## 4. Experimental Result and Discussion

### 4.1 Multi-centered datasets and preprocessing

We include ADNI [48] (ADNI1, ADNI2, and ADNI Go), AIBL [49] and OASIS3 [50] datasets to validate the proposed method for AD diagnosis and staging prediction. Detailed demographic information is shown in Table 2.

1) ADNI: The primary objective of the ADNI has been to investigate how the use of MRI scans [51], along with various biological markers, can enhance our understanding of the stages and progression of MCI and AD. Our analysis includes all patients who underwent MRI and clinical factors. This resulted in a total of 774 patients and 3880 MRIs.

2) AIBL: The Australian Imaging, Biomarkers and Lifestyle (AIBL) [49] study is a longitudinal research project designed to investigate AD and MCI through comprehensive follow-up of participants over time. To address data imbalance and ensure the inclusion of subjects with both MRI scans and corresponding diagnostic labels, we selected 443 participants with a total of 642 sMRI sessions.

3) OASIS3: OASIS3 [50] includes 1,474 participants from various racial and ethnic backgrounds, with 84% identified as Caucasian and 15% as African American, along with five individuals reporting Hispanic ethnicity. To construct a balanced dataset, all available AD and MCI sessions were included, while a random subset of 371 CN sessions was sampled for comparison.

**Table 2**. Demographic information for the enrolled datasets.

| Datasets | ADNI | AIBL | OASIS3 |
|---|---|---|---|
| Number of sMRIs | CN: 1663; MCI: 1244; AD: 972 | CN: 300; MCI: 185; AD: 157 | CN: 371; MCI: 83; AD: 332 |
| Avg. # MRIs/ Subject | 5.22 | 1.45 | 1.26 |
| Sex (M/F) | 430 / 314 | 203 / 239 | 172 / 362 |
| Ages | 76.07 ± 6.93 | 74.14 ± 7.05 | 72.47 ± 8.37 |
| CDR score | 0.39 ± 0.45 | 0.37 ± 0.55 | 0.32 ± 0.36 |
| MMSE score | 26.64 ± 3.92 | 26.05 ± 5.11 | 27.48 ± 3.19 |

### 4.2. Implementation Details and Experimental Setup

We implemented the proposed GMN4AD-TTA using PyTorch 1.13 and the grid search was performed to fine-tune the optimal parameters. For each dataset, we applied the stratified sampling to select 10% subjects as the templates and the rest we performed a 5-fold cross validation. The Adam optimizer, with an initial learning rate of 0.0001 was employed. We used the exponential decay strategy in the training phase to adjust the learning rate, and we set the decay rate as 0.99 for each 20 training epochs. The hyperparameters, including the number of intra-graph feature embedding (2, 3, 4, 5), the number of cross-graph feature embedding (3, 4, 5) and the number of hidden units (64, 96, 128, 192) were adjusted using a grid search. For each dataset and task, we selected 10% samples as the template set and the rest, we conducted a 5-fold cross validation. Best model parameters are shown in Table S1 in the supplementary materials.

### 4.3. AD diagnosis performance

We apply our GMN4AD model on three datasets, including ADNI, AIBL and OASIS individually on these three tests for AD vs. CN, AD vs. MCI, and MCI vs. CN classification tasks. Table 3 shows the classification results of the proposed GMN4AD.

**Table 3**. Performance of the AD detection task using the proposed GMN4AD.

| Dataset | Task | ACC | PREC | F1 | SEN | SPEC |
|---|---|---|---|---|---|---|
| ADNI | CN | 0.9237 ± 0.0427 | 0.9036 ± 0.0173 | 0.8902 ± 0.0677 | 0.8826 ± 0.1146 | 0.9474 ± 0.0061 |
| AIBL | vs. | 0.9258 ± 0.0294 | 0.9230 ± 0.0403 | 0.8774 ± 0.0419 | 0.8371 ± 0.0544 | 0.9673 ± 0.0191 |
| OASIS3 | AD | 0.7803 ± 0.0227 | 0.7382 ± 0.0362 | 0.7796 ± 0.0306 | 0.8272 ± 0.0436 | 0.7359 ± 0.0459 |
| ADNI | MCI | 0.7613 ± 0.0257 | 0.7165 ± 0.0425 | 0.7317 ± 0.0228 | 0.7516 ± 0.0533 | 0.7657 ± 0.0708 |
| AIBL | vs. | 0.7328 ± 0.0137 | 0.7484 ± 0.0592 | 0.7114 ± 0.0075 | 0.6829 ± 0.0501 | 0.7856 ± 0.0385 |
| OASIS3 | AD | 0.8108 ± 0.0331 | 0.8089 ± 0.0322 | 0.8941 ± 0.0198 | 1.0000 ± 0.0000 | 0.0568 ± 0.0634 |
| ADNI | CN | 0.8112 ± 0.0193 | 0.7995 ± 0.0404 | 0.7761 ± 0.0195 | 0.7568 ± 0.0441 | 0.8536 ± 0.0372 |
| AIBL | vs. | 0.7732 ± 0.0372 | 0.7603 ± 0.1231 | 0.6422 ± 0.0491 | 0.5655 ± 0.0602 | 0.8914 ± 0.0668 |
| OASIS3 | MCI | 0.8350 ± 0.0691 | 0.5000 ± 0.1920 | 0.3172 ± 0.1623 | 0.2467 ± 0.1543 | 0.9571 ± 0.0194 |

Table 3 summarizes the classification performance of GMN4AD across the ADNI, AIBL, and OASIS3 datasets under three diagnostic settings. In the AD vs. CN task, ADNI and AIBL reach accuracies above 0.92 with balanced precision, recall, and specificity, while OASIS3 attains 0.78 accuracy and 0.83 sensitivity, showing greater variation across centers. For the AD vs. MCI comparison, the accuracy ranges from 0.73 to 0.81, and the F1 scores remain above 0.70 on ADNI and AIBL, suggesting moderate separability between the two stages. In the MCI vs. CN task, all datasets yield accuracies between 0.77 and 0.83, with specificity consistently higher than sensitivity, indicating that most cognitively normal cases are correctly identified while mild impairment is harder to capture.

Across all datasets, ADNI generally shows the most stable results, followed by AIBL, whereas OASIS3 exhibits larger fluctuations in sensitivity and specificity. These observations indicate that GMN4AD maintains comparable performance across different cohorts and diagnostic boundaries, and that dataset composition and imaging variability play a measurable role in classification outcomes.

In addition, we compared the proposed GMN4AD with the several state-of-the-art approaches for AD classification using graph and structured MRIs, including:

- Graph convolution network (GCN) [52]: In this study, a Graph Convolutional Network (GCN) was applied to MRI-derived graphs, using ROI-to-ROI similarity as the adjacency matrix and voxel-level features for each ROI.
- Graph attention network (GAT) [53]: This paper employs a Graph Attention Network (GAT) to classify Alzheimer's disease, mild cognitive impairment, and healthy controls by learning attention-weighted relationships between subjects from neuroimaging-derived features.
- Graph isomorphism network (GIN) [54]: DMT-GIN builds brain-connectivity graphs from fMRI, uses a GIN with attention and multi-task heads (AD diagnosis plus age/gender auxiliary tasks) to improve AD vs. MCI vs. CN classification.
- Feature inductive learning (FIL) [22]: FIL address the heterogeneity of the brain imaging data by mapping all features into a shared latent feature space and the graph model is applied for AD diagnosis.
- Baseline (BL): In our GMN4AD model, we incorporated cross-graph feature representation learning prior to the graph matching stage. In the baseline model, this submodule was removed to evaluate the effectiveness of the cross-graph feature embedding.

We implemented the above methods and fit their models into our AD classification tasks. For each model, we conducted the grid search and performed the 5-fold cross validation. The results on the ADNI datasets are shown in Table 4 and the comparison of AIBL and OASIS3 datasets are shown in Tables S2 and S3 in the supplementary materials.

**Table 4. Comparative experiment results on ADNI dataset**

| Task | model | ACC | PREC | F1 | SEN | SPEC |
|---|---|---|---|---|---|---|
| AD | GCN | 0.8773±0.0133 | 0.8512±0.0212 | 0.8282±0.0221 | 0.8069±0.0317 | 0.9184±0.0084 |
| vs. | GAT | 0.8769±0.0128 | 0.8523±0.0210 | 0.8276±0.0186 | 0.8049±0.0289 | 0.9190±0.0112 |
| CN | GIN | 0.8765±0.0140 | 0.8489±0.0366 | 0.8282±0.0169 | 0.8098±0.0252 | 0.9158±0.0230 |
| | FIL | 0.8878±0.0095 | 0.8529±0.0254 | 0.8461±0.0145 | 0.8398±0.0165 | 0.9157±0.0155 |
| | BL | 0.8429±0.0211 | 0.7705±0.0435 | 0.7898±0.0268 | 0.8140±0.0552 | 0.8595±0.0394 |
| | ours | **0.9237±0.0427** | **0.9036±0.0173** | **0.8902±0.0676** | **0.8826±0.1146** | **0.9474±0.0061** |
| AD vs. MCI | GCN | 0.7609±0.0165 | **0.7207±0.0368** | 0.7276±0.0233 | 0.7391±0.0602 | 0.7778±0.0499 |
| | GAT | 0.7549±0.0142 | 0.7105±0.0500 | 0.7238±0.0141 | 0.7413±0.0316 | 0.7661±0.0489 |
| | GIN | 0.7544±0.0184 | 0.7193±0.0318 | 0.7150±0.0242 | 0.7113±0.0273 | **0.7865±0.0290** |
| | FIL | 0.7529±0.0331 | 0.7153±0.0213 | 0.7164±0.0318 | 0.7195±0.0586 | 0.7789±0.0362 |
| | BL | 0.6299±0.0994 | 0.5861±0.1031 | 0.5935±0.1117 | 0.6420±0.1716 | 0.6225±0.2230 |
| | ours | **0.7613±0.0257** | 0.7165±0.0425 | **0.7317±0.0228** | **0.7516±0.0533** | 0.7657±0.0708 |
| CN vs. MCI | GCN | 0.7983±0.0168 | 0.7690±0.0308 | 0.7649±0.0324 | 0.7629±0.0542 | 0.8248±0.0274 |
| | GAT | 0.8025±0.0134 | 0.7832±0.0258 | 0.7670±0.0192 | 0.7528±0.0362 | 0.8398±0.0274 |
| | GIN | 0.8002±0.0111 | 0.7763±0.0276 | 0.7663±0.0132 | 0.7574±0.0234 | 0.8338±0.0166 |
| | FIL | 0.7983±0.0207 | 0.7676±0.0398 | 0.7670±0.0243 | **0.7675±0.0259** | 0.8225±0.0299 |
| | BL | 0.6636±0.0480 | 0.5974±0.0716 | 0.6276±0.0880 | 0.6721±0.1198 | 0.6573±0.0898 |
| | ours | **0.8112±0.0193** | **0.7995±0.0404** | **0.7761±0.0195** | 0.7568±0.0441 | **0.8536±0.0372** |

Table 4 presents the comparative results of GMN4AD and several representative graph-based models on the ADNI dataset. Across all three diagnostic tasks, GMN4AD consistently achieves the highest or comparable scores among all methods. In the AD vs. CN task, it attains 0.9237 accuracy and 0.9474 specificity, exceeding GCN, GAT, GIN, and FIL by roughly 3–5%, indicating that the proposed framework maintains a strong balance between sensitivity and specificity. For the AD vs. MCI task, the performance remains stable with 0.7613 accuracy, slightly outperforming other methods ($\approx$0.75) and maintaining comparable F1 values. In the CN vs. MCI setting, GMN4AD yields 0.8112 accuracy and the highest specificity (0.8536), suggesting reliable detection of cognitively normal individuals while preserving consistent precision across folds. These numerical results demonstrate that GMN4AD sustains steady performance across varying diagnostic boundaries within the same dataset.

Compared with single-graph architectures such as GCN or GAT, GMN4AD incorporates an iterative cross-graph matching framework designed to explicitly capture structural correspondences across subjects' brain graphs. This mechanism allows the model to integrate complementary information from paired graphs, maintaining stable discriminative performance across different classification settings.

### 4.4. Performance of Test-time Domain Adaptation

To improve the inference accuracy across different datasets, we applied our contrastive learning based TTA to GMN4AD models. The TTA pipeline fine-tunes the pre-trained graph matching model on unlabeled target data to adapt it to domain-specific characteristics. The adaptation training set includes a mix of source and target samples. A TTA-specific optimizer, i.e. Adam, is defined to update only selective layers of the model, including graph convolution and cross-graph modules, while keeping other parameters fixed. During each adaptation epoch, dropout is disabled and the model is iteratively optimized using a $\mathcal{L}_{\text{TTA}}$ defined in Eq. 8. on target data.

To demonstrate the improvement of the proposed method, we compared the TTA results with other peer methods, including:

- Source-only inference (**SOI**). In this baseline, the model trained on the source dataset is directly applied to the target dataset without any adaptation.
- Self-supervised **D**omain **A**daptation with **G**raph **E**mbedding (**DAGE**) [37]: DAGE leverages self-supervised learning to align source and target graph embeddings by reconstructing the adjacency matrix according to the graph embedding, enabling cross-domain generalization without requiring labeled target data.
- **M**arginal **E**ntropy **M**inimization with **O**ne test point (**MEMO**) [32]: MEMO adapts a pretrained model at test time by minimizing the entropy of its predictions over augmented versions of a single test input, encouraging confident and invariant predictions without using labels.

We applied the aforementioned methods to our AD classification tasks by integrating their models with the feature embeddings derived from graph matching. For each approach, a grid search was conducted followed by a five-fold cross-validation to optimize performance. The experimental results on the ADNI dataset are presented in Table 4, while comparisons on the AIBL and OASIS3 datasets are provided in Tables S4 and S5 of the supplementary materials.

**Table 5**. Comparative experiment results on ADNI dataset as source dataset and prediction on target dataset using unsupervised domain adaptation.

| Target Dataset | Task | Method | ACC | PREC | F1 | SEN | SPEC |
|---|---|---|---|---|---|---|---|
| AIBL | AD vs. CN | SOI | 0.8261±0.0397 | 0.7446±0.0931 | 0.7473±0.0664 | 0.7610±0.0951 | 0.8608±0.0587 |
| | | DAGE | 0.6823±0.1205 | 0.5496±0.1718 | 0.6463±0.1127 | **0.8231±0.1048** | 0.6099±0.2062 |
| | | MEMO | 0.7521±0.1359 | 0.7304±0.1864 | 0.6553±0.0803 | 0.6520±0.1725 | 0.8025±0.2835 |
| | | Ours | **0.8586±0.0232** | **0.8190±0.0428** | **0.7830±0.0447** | 0.7570±0.0997 | **0.9074±0.0427** |
| | AD vs. MCI | SOI | 0.5899±0.0383 | 0.5522±0.0514 | 0.6501±0.0392 | **0.8012±0.0891** | 0.4004±0.0678 |
| | | DAGE | **0.6600±0.0577** | **0.6633±0.0727** | 0.6411±0.0171 | 0.6306±0.0653 | **0.6750±0.1616** |
| | | MEMO | 0.6212±0.0278 | 0.5734±0.0454 | 0.6709±0.0556 | 0.8203±0.1278 | 0.4165±0.1781 |
| | | Ours | 0.6443±0.0462 | 0.6465±0.0580 | **0.6843±0.0495** | 0.7182±0.0139 | 0.5851±0.1317 |
| | CN vs. MCI | SOI | 0.6450±0.0317 | 0.4954±0.0622 | 0.4931±0.0872 | 0.4940±0.1155 | 0.7239±0.0449 |
| | | DAGE | 0.5937±0.1139 | 0.4836±0.1229 | 0.4860±0.0383 | 0.5389±0.1374 | 0.6265±0.2540 |
| | | MEMO | 0.5367±0.1316 | 0.4404±0.1095 | 0.4862±0.0700 | **0.6318±0.2258** | 0.5001±0.2870 |
| | | Ours | **0.6695±0.0523** | **0.5566±0.1152** | **0.5396±0.0776** | 0.5546±0.1349 | **0.7369±0.1216** |
| OASIS | AD vs. CN | SOI | 0.7464±0.0313 | **0.8195±0.0459** | 0.6860±0.0276 | 0.5927±0.0476 | 0.8845±0.0324 |
| | | DAGE | 0.5743±0.0977 | 0.5497±0.1007 | 0.6085±0.0601 | 0.7023±0.0976 | 0.4660±0.2088 |
| | | MEMO | 0.6714±0.0542 | 0.6589±0.1236 | 0.6734±0.0224 | **0.7329±0.1498** | 0.6102±0.2609 |
| | | Ours | **0.7788±0.0315** | 0.8155±0.0472 | **0.7447±0.0241** | 0.6902±0.0618 | **0.8576±0.0505** |
| | AD vs. MCI | SOI | 0.5229±0.0619 | 0.8525±0.0952 | 0.6171±0.0748 | 0.4904±0.0869 | 0.6682±0.2361 |
| | | DAGE | 0.5600±0.0752 | 0.8585±0.0531 | 0.6573±0.0775 | 0.5390±0.0989 | 0.6553±0.1138 |
| | | MEMO | **0.6800±0.1137** | 0.8353±0.0644 | **0.7783±0.1021** | **0.7381±0.1472** | 0.4378±0.2018 |
| | | Ours | 0.5743±0.0445 | **0.8823±0.0595** | 0.6674±0.0398 | 0.5371±0.0336 | **0.7199±0.1188** |
| | CN vs. MCI | SOI | 0.7531±0.0589 | 0.3477±0.1259 | 0.3516±0.0398 | 0.4152±0.1357 | 0.8215±0.1013 |
| | | DAGE | 0.6840±0.0771 | 0.2701±0.0449 | 0.3384±0.0413 | 0.4875±0.1196 | 0.7215±0.1040 |
| | | MEMO | 0.6415±0.2233 | 0.3035±0.1116 | 0.3757±0.0927 | **0.5922±0.2149** | 0.6531±0.2927 |
| | | Ours | **0.8036±0.0323** | **0.4283±0.0880** | **0.4560±0.0631** | 0.4995±0.0901 | **0.8626±0.0525** |

As shown in Table 5, the proposed TTA strategy consistently enhances the cross-dataset generalization of GMN4AD compared with the SOI baseline. For the AD vs. CN task, our model achieves the highest or near-highest accuracy on both AIBL and OASIS3 target domains (0.8586 and 0.7788), with balanced sensitivity and specificity, demonstrating that the adaptation procedure effectively preserves structural discriminability under cross-domain shifts. In the AD vs. MCI classification, although the performance gains are less pronounced, GMN4AD with TTA remains competitive.

For the AD vs. MCI classification task, on AIBL, DAGE attains slightly higher accuracy (0.6600 vs 0.6443), whereas our model achieves superior F1 (0.6843 vs 0.6411), reflecting more balanced sensitivity–precision trade-offs. On OASIS3, MEMO records higher accuracy and F1 (0.6800 vs 0.5743; 0.7783 vs 0.6674), yet our method yields greater precision (0.8823 vs 0.8353) and specificity (0.7199 vs 0.4378), suggesting more reliable and conservative decision boundaries under severe domain shifts. Additionally, for the CN–MCI task, GMN4AD exhibits steady improvements across both target domains, underscoring its robustness in modeling mild structural variations near diagnostic boundaries.

Overall, while other methods may surpass ours in isolated metrics, the proposed TTA-enhanced GMN4AD achieves the most stable and balanced performance across datasets, demonstrating strong resilience to domain shifts, particularly in the early and transitional stages of Alzheimer's disease.

### 4.5. Model interpretation for AD diagnosis

**Feature importance**. To interpret the contribution of individual radiomic features to the graph matching–based AD diagnosis, we applied the ZORRO-Paired explainer across all graph pairs. Our handcrafted feature set includes 113 descriptors capturing pixel intensity distributions, spatial positioning, and texture characteristics derived from MRI scans. Using a fidelity threshold of $\tau = 0.99$ and selecting the top-1 feature per graph pair, we identified a consistent subset of features that significantly influenced similarity predictions. The top-15 most important features for these three tasks using ADNI dataset are shown in Figure 2.

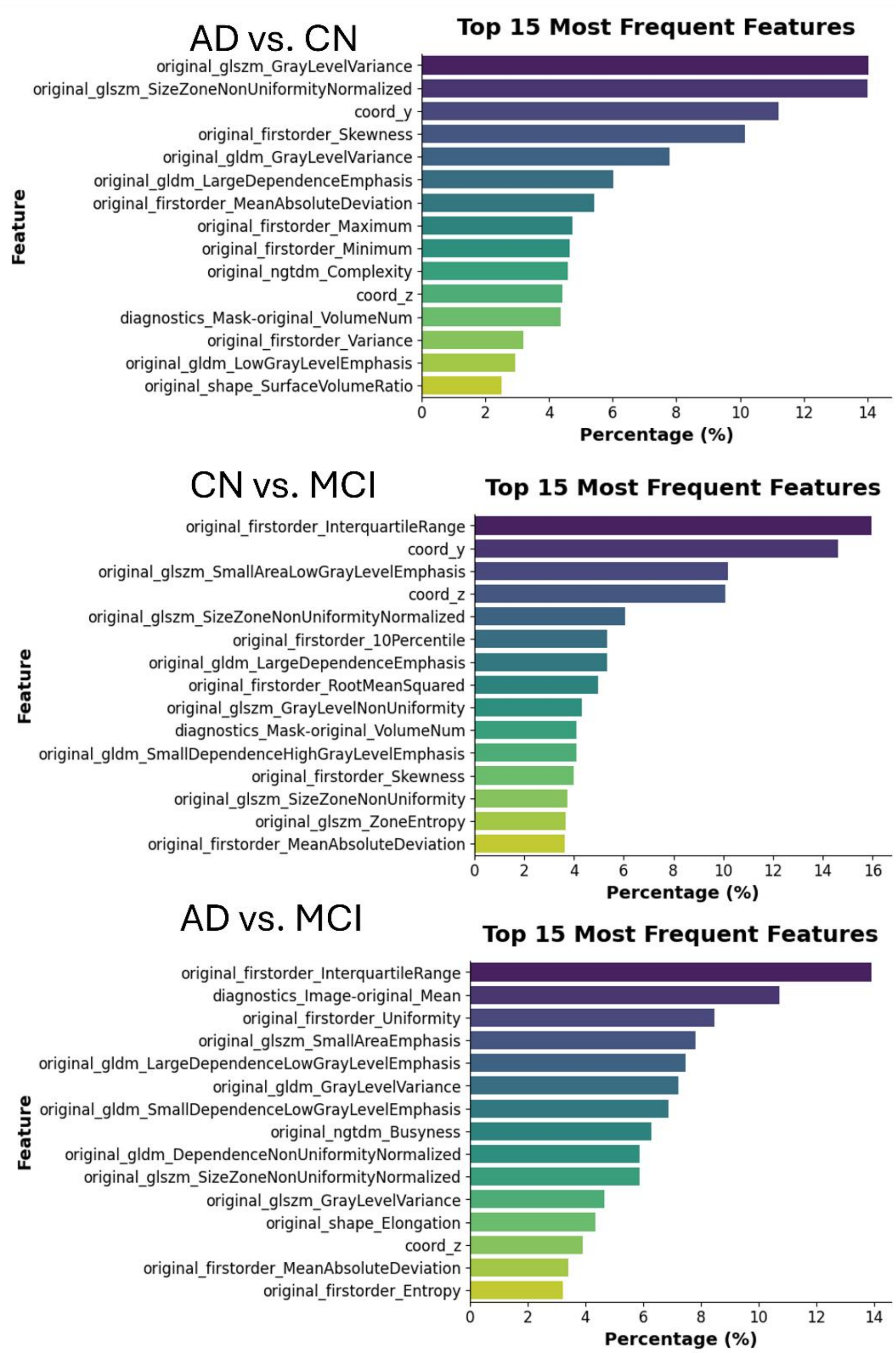

**Figure 2**. Radiomics Feature Importance Across Disease Comparisons. The bar plots display the top features contributing to disease classification in ADNI (AD vs CN, CN vs MCI, and AD vs MCI). Each bar represents a radiomics feature extracted from MRI, including first-order intensity, texture (GLSZM, GLDM, NGTDM), shape, and coordinate-based features (coord_x, coord_y, coord_z), with the horizontal axis showing the relative importance as a percentage of times the feature was selected in the ZORRO-Paired explanation

The feature importance analysis reveals that texture and first-order intensity features are consistently the most informative across comparisons, highlighting subtle microstructural changes in brain tissue associated with AD progression. In the AD vs. CN task, features such as GLSZM_GrayLevelVariance, SizeZoneNonUniformityNormalized, and firstorder_Skewness dominate, reflecting heterogeneity in voxel intensities and overall intensity distribution shifts in key brain regions. Coordinate features (coord_y and coord_z) also rank highly, indicating positional changes in brain structures such as ventricular enlargement and hippocampal atrophy, which are known hallmarks of AD. Similarly, in the CN vs. MCI comparison, intensity variability features (e.g., firstorder_InterquartileRange) and small-area texture measures capture early microstructural changes, while coordinate shifts point to subtle anatomical displacements occurring in mild cognitive impairment.

In the AD vs. MCI comparison, first-order and texture features remain dominant, including InterquartileRange, Uniformity, SmallAreaEmphasis, and GLDM_LargeDependenceLowGrayLevelEmphasis. These features capture progressive tissue heterogeneity, intensity alterations, and structural organization changes, while the presence of coord_z highlights superior-inferior shifts in key regions such as the hippocampus and ventricles. Overall, the analysis shows that both microstructural heterogeneity and macroscopic positional changes are critical for distinguishing disease stages, with texture and intensity features reflecting tissue pathology and coordinate features reflecting structural shifts due to atrophy. This aligns with the known neuropathology of AD, where white matter integrity loss, hippocampal atrophy, and ventricular enlargement are closely associated with cognitive decline and disease progression.

**ROI importance**. To further localize the model's decision-making process, we analyzed node-level importance scores generated by ZORRO-Paired. Similar to feature importance analysis, ZORRO-Paired begins by randomly masking all brain ROIs and progressively adds the most informative ROIs for graph matching. Once the fidelity score reaches the pre-defined threshold, the explanation process is stopped, and the frequency of each included brain ROI is recorded. A higher frequency indicates greater importance. Each node corresponds to a specific brain region segmented from MRI data, and the explainer identifies which regions most strongly contribute to graph similarity predictions. In Figure 3, we visualize the top 10 most important ROIs identified by GMM4AD across different datasets and tasks.

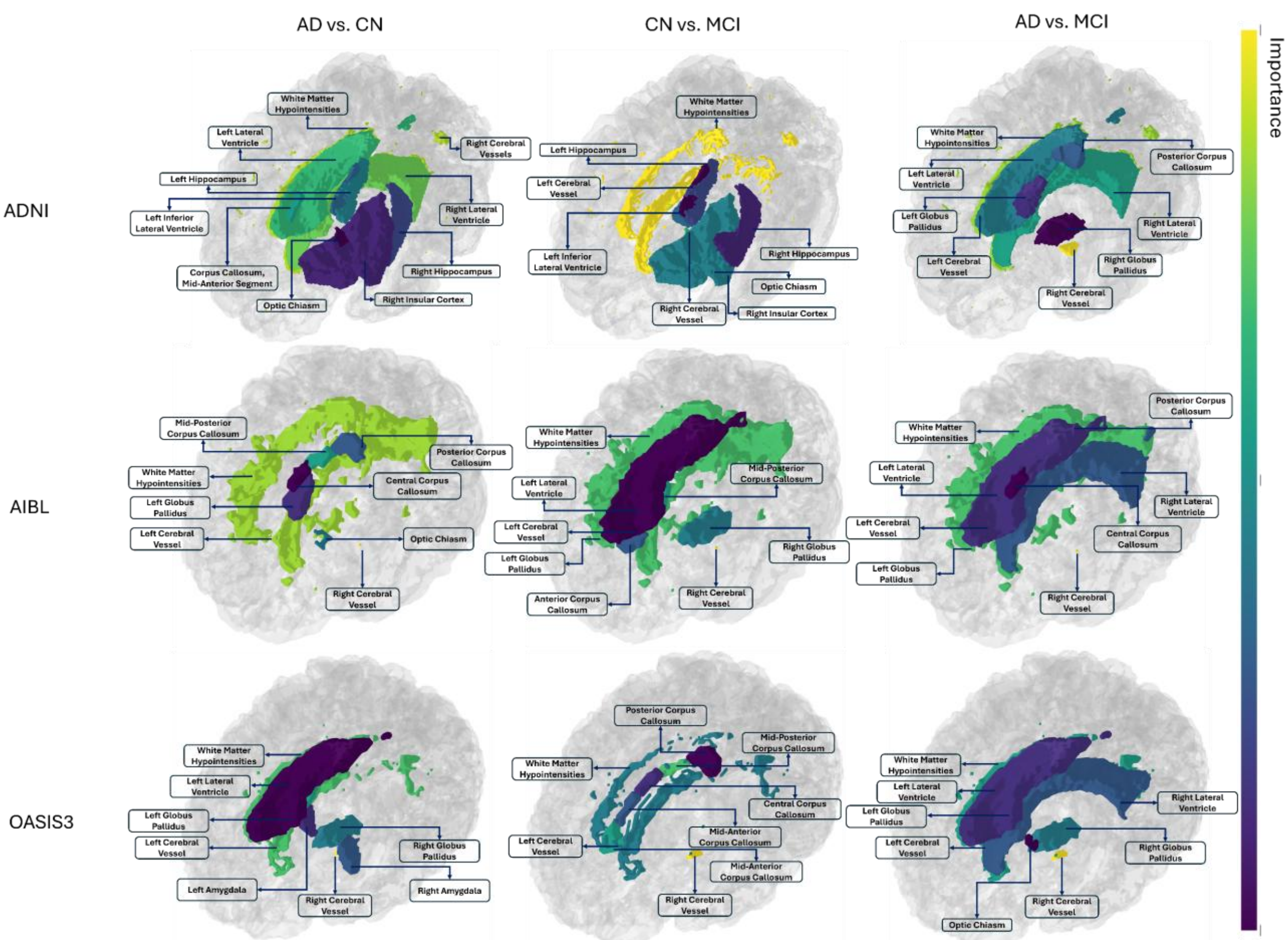

**Figure 3**. Top 10 most important brain ROIs identified by GMM4AD. The figure visualizes the ROIs that contribute most strongly to the model's graph similarity predictions across different datasets and tasks.

In Figure 3, each ROI is colored according to its relative importance score, with brighter colors (e.g., light yellow) indicating higher importance and darker colors (e.g., violet) indicating lower importance. ROIs are ranked based on their frequency of inclusion in the ZORRO-Paired explanation process, reflecting how strongly each region influences the model's predictions. If fewer than 10 ROIs are annotated, it indicates that the left and right cerebral white matter are highly important, although they are not visualized due to their large volume, which would obscure smaller regions. White matter changes, such as demyelination and small vessel disease, are known to occur early in Alzheimer's disease and can disrupt brain connectivity. Their high importance in our model is consistent with prior findings linking cerebral white matter integrity to cognitive decline and disease progression, emphasizing their critical role in AD diagnosis.

Among all datasets and comparisons, three brain regions consistently emerged as the most important ROIs: Right Cerebral Vessel, Left Cerebral Vessel, and White Matter Hypointensities. These regions are closely associated with vascular health and cerebral perfusion. Abnormalities in cerebral vessels and white matter, including hypointensities [55], are well-known markers of small vessel disease and microvascular pathology [56], which have been linked to cognitive decline and AD progression. The consistent importance of these ROIs across AD vs. CN, CN vs. MCI, and AD vs. MCI comparisons highlights the critical role of vascular and white matter integrity in AD diagnosis and underscores the value of incorporating vascular and white matter features in predictive models for early detection.

**5. Limitation and future work**

A current limitation of the proposed GMN4AD framework is that, during inference, graph matching must be conducted between each test subject and all subjects in the template set, which can be computationally demanding. Reducing the number of subjects in the template set would shorten prediction time but may risk performance degradation. To address this, future work will explore graph clustering techniques to identify representative subjects within each cluster, allowing for a more compact yet informative template set that accelerates prediction without compromising accuracy.

Additionally, the current approach is limited to binary graph similarity matching, which allows only pairwise distinctions such as AD vs. CN, AD vs. MCI, or CN vs. MCI. In future work, the graph matching framework should be extended to support multi-class similarity learning, enabling simultaneous differentiation among all diagnostic categories.

Furthermore, the present model relies solely on MRI, capturing structural brain anatomy. Incorporating Positron Emission Tomography (PET) imaging, which provides complementary functional information [57], and developing multi-view graph matching techniques will allow the model to leverage both structural and functional modalities for a more comprehensive assessment of brain alterations in Alzheimer's disease.,

**6. Conclusion**

In this paper, we proposed a graph matching based method for AD diagnosis using pairwise brain-derived graphs. By measuring the similarity between the testing brain graph and template brain graph, GMN4AD assesses the similarity between them and assigns the AD diagnosis prediction results according to the label of the template graph. During the test phase for domain adaptation, we proposed the hybrid test-time domain adaptation strategy, consisting of self-contrastive learning and graph reconstruction for unsupervised domain adaptation techniques and adapted the trained GMN4AD from source domain to target domain for robust generality. Experimental results demonstrate the superior performance of the proposed GMN4AD in identifying AD vs. CN, AD vs. MCI and MCI vs. CN.

**Acknowledgement:**

This work is partially supported by American Heart Association under award #25AIREA1377168 (PI. Chen Zhao). We gratefully acknowledge the ADNI, AIBL, and OASIS teams for providing access to their valuable imaging datasets, which made this study possible.

**Reference**

1. Gonzales, M. M. *et al.* Biological aging processes underlying cognitive decline and neurodegenerative disease. *J. Clin. Invest.* **132**, e158453 (2022).
2. Matthews, K. A. *et al.* Racial and ethnic estimates of Alzheimer's disease and related dementias in the United States (2015–2060) in adults aged ≥65 years. *Alzheimers Dement.* **15**, 17–24 (2019).
3. Mitka, M. PET Imaging for Alzheimer Disease: Are Its Benefits Worth the Cost? *JAMA* **309**, 1099 (2013).
4. Veitch, D. P. *et al.* Understanding disease progression and improving Alzheimer's disease clinical trials: Recent highlights from the Alzheimer's Disease NeuRoImaging Initiative. *Alzheimers Dement.* **15**, 106–152 (2019).
5. Kikuchi, M. *et al.* Identification of mild cognitive impairment subtypes predicting conversion to Alzheimer's disease using multimodal data. *Comput. Struct. Biotechnol. J.* **20**, 5296–5308 (2022).
6. Petersen, R. C. Mild cognitive impairment as a diagnostic entity. *J. Intern. Med.* **256**, 183–194 (2004).
7. Silva-Spínola, A., Baldeiras, I., Arrais, J. P. & Santana, I. The Road to Personalized Medicine in Alzheimer's Disease: The Use of Artificial Intelligence. *Biomedicines* **10**, 315 (2022).
8. Frisoni, G. B., Fox, N. C., Jack, C. R., Scheltens, P. & Thompson, P. M. The clinical use of structural MRI in Alzheimer disease. *Nat. Rev. Neurol.* **6**, 67–77 (2010).
9. Dukart, J. *et al.* Relationship between imaging biomarkers, age, progression and symptom severity in Alzheimer's disease. *NeuRoImage Clin.* **3**, 84–94 (2013).
10. Bloudek, L. M., Spackman, D. E., Blankenburg, M. & Sullivan, S. D. Review and Meta-Analysis of Biomarkers and Diagnostic Imaging in Alzheimer's Disease. *J. Alzheimers Dis.* **26**, 627–645 (2011).
11. Zhang, Y. *et al.* A novel spatiotemporal graph convolutional network framework for functional connectivity biomarkers identification of Alzheimer's disease. *Alzheimers Res. Ther.* **16**, 60 (2024).
12. Lei, B. *et al.* Alzheimer's disease diagnosis from multi-modal data via feature inductive learning and dual multilevel graph neural network. *Med. Image Anal.* **97**, 103213 (2024).
13. Avelar-Pereira, B., Belloy, M. E., O'Hara, R., Hosseini, S. M. H., & for the Alzheimer's Disease NeuRoImaging Initiative. Decoding the heterogeneity of Alzheimer's disease diagnosis and progression using multilayer networks. *Mol. Psychiatry* **28**, 2423–2432 (2023).
14. Singh, A., Shi, W. & Wang, M. D. Multi-Modal Deep Feature Integration for Alzheimer's Disease Staging. in *2023 IEEE International Conference on Bioinformatics and Biomedicine (BIBM)* 1–6 (IEEE, Istanbul, Turkiye, 2023). doi:10.1109/BIBM58861.2023.10431906.
15. Venugopalan, J., Tong, L., Hassanzadeh, H. R. & Wang, M. D. Multimodal deep learning models for early detection of Alzheimer's disease stage. *Sci. Rep.* **11**, 1–13 (2021).
16. Lei, B. *et al.* Federated Domain Adaptation via Transformer for Multi-Site Alzheimer's Disease Diagnosis. *IEEE Trans. Med. Imaging* **42**, 3651–3664 (2023).
17. Khemani, B., Patil, S., Kotecha, K. & Tanwar, S. A review of graph neural networks: concepts, architectures, techniques, challenges, datasets, applications, and future directions. *J. Big Data* **11**, 18 (2024).
18. Khojaste-Sarakhsi, M., Haghighi, S. S., Ghomi, S. M. T. F. & Marchiori, E. Deep learning for Alzheimer's disease diagnosis: A survey. *Artif. Intell. Med.* **130**, 102332 (2022).
19. Li, J. *et al.* Dual Attention Graph Convolutional Network Fusing Imaging and Genetic Data for Early Alzheimer's Disease Diagnosis. in *2024 46th Annual International Conference of the IEEE Engineering in Medicine and Biology Society (EMBC)* 1–4 (IEEE, Orlando, FL, USA, 2024). doi:10.1109/EMBC53108.2024.10781972.
20. Zhang, Y., He, X., Chan, Y. H., Teng, Q. & Rajapakse, J. C. Multi-modal graph neural network for early diagnosis of Alzheimer's disease from sMRI and PET scans. *Comput. Biol. Med.* **164**, 107328 (2023).
21. Yang, F. *et al.* Multi-model adaptive fusion-based graph network for Alzheimer's disease prediction. *Comput. Biol. Med.* **153**, 106518 (2023).

22. Lei, B. *et al.* Alzheimer's disease diagnosis from multi-modal data via feature inductive learning and dual multilevel graph neural network. *Med. Image Anal.* **97**, 103213 (2024).

23. Qu, G., Zhou, Z., Calhoun, V. D., Zhang, A. & Wang, Y.-P. Integrated brain connectivity analysis with fMRI, DTI, and sMRI powered by interpretable graph neural networks. *Med. Image Anal.* **103**, 103570 (2025).

24. Zhou, H., He, L., Zhang, Y., Shen, L. & Chen, B. Interpretable Graph Convolutional Network Of Multi-Modality Brain Imaging For Alzheimer's Disease Diagnosis. in *2022 IEEE 19th International Symposium on Biomedical Imaging (ISBI)* 1–5 (IEEE, Kolkata, India, 2022). doi:10.1109/ISBI52829.2022.9761449.

25. Lei, B. *et al.* Adaptive sparse learning using multi-template for neurodegenerative disease diagnosis. *Med. Image Anal.* **61**, 101632 (2020).

26. Yan, J. *et al.* A Short Survey of Recent Advances in Graph Matching. in *Proceedings of the 2016 ACM on International Conference on Multimedia Retrieval* 167–174 (ACM, New York New York USA, 2016). doi:10.1145/2911996.2912035.

27. Ma, G., Ahmed, N. K., Willke, T. L. & Yu, P. S. Deep graph similarity learning: a survey. *Data Min. Knowl. Discov.* **35**, 688–725 (2021).

28. Gu, Y. *et al.* Structure-aware siamese graph neural networks for encounter-level patient similarity learning. *J. Biomed. Inform.* **127**, 104027 (2022).

29. Zhao, C., Esposito, M., Xu, Z. & Zhou, W. HAGMN-UQ: Hyper Association Graph Matching Network with Uncertainty Quantification for Coronary Artery Semantic Labeling. *Med. Image Anal.* 103374 (2024) doi:10.1016/j.media.2024.103374.

30. Liang, J., He, R. & Tan, T. A Comprehensive Survey on Test-Time Adaptation Under Distribution Shifts. *Int. J. Comput. Vis.* **133**, 31–64 (2025).

31. Kundu, J. N. *et al.* Concurrent Subsidiary Supervision for Unsupervised Source-Free Domain Adaptation. in *Computer Vision – ECCV 2022* (eds Avidan, S., Brostow, G., Cissé, M., Farinella, G. M. & Hassner, T.) vol. 13690 177–194 (Springer Nature Switzerland, Cham, 2022).

32. Zhang, M., Levine, S. & Finn, C. MEMO: Test Time Robustness via Adaptation and Augmentation. in *Advances in Neural Information Processing Systems* (eds Koyejo, S. et al.) vol. 35 38629–38642 (Curran Associates, Inc., 2022).

33. Nado, Z. *et al.* Evaluating Prediction-Time Batch Normalization for Robustness under Covariate Shift. Preprint at https://doi.org/10.48550/arXiv.2006.10963 (2021).

34. Zhang, J., Qi, L., Shi, Y. & Gao, Y. DomainAdaptor: A Novel Approach to Test-time Adaptation. in 18971–18981 (2023).

35. Karani, N., Erdil, E., Chaitanya, K. & Konukoglu, E. Test-time adaptable neural networks for robust medical image segmentation. *Med. Image Anal.* **68**, 101907 (2021).

36. Shu, M. *et al.* Test-Time Prompt Tuning for Zero-Shot Generalization in Vision-Language Models. *Adv. Neural Inf. Process. Syst.* **35**, 14274–14289 (2022).

37. Hedegaard, L., Sheikh-Omar, O. A. & Iosifidis, A. Supervised Domain Adaptation: A Graph Embedding Perspective and a Rectified Experimental Protocol. *IEEE Trans. Image Process.* **30**, 8619–8631 (2021).

38. Fischl, B. FreeSurfer. *NeuRoImage* **62**, 774–781 (2012).

39. Zhao, C. *et al.* AGMN: Association graph-based graph matching network for coronary artery semantic labeling on invasive coronary angiograms. *Pattern Recognit.* **143**, 109789 (2023).

40. Zhao, C. *et al.* Lung segmentation and automatic detection of COVID-19 using radiomic features from chest CT images. *Pattern Recognit.* **119**, 108071 (2021).

41. Adams, R. P. & Zemel, R. S. Ranking via Sinkhorn Propagation. Preprint at http://arxiv.org/abs/1106.1925 (2011).

42. Cruz, R. S., Fernando, B., Cherian, A. & Gould, S. Visual Permutation Learning. *IEEE Trans. Pattern Anal. Mach. Intell.* **41**, 3100–3114 (2019).

43. Neural Graph Matching for Pre-training Graph Neural Networks. in *Proceedings of the 2022 SIAM International Conference on Data Mining (SDM)* (eds Hou, Y. et al.) (Society for Industrial and Applied Mathematics, Philadelphia, PA, 2022). doi:10.1137/1.9781611977172.20.

44. Wang, R., Yan, J. & Yang, X. Combinatorial Learning of Robust Deep Graph Matching: an Embedding based Approach. *IEEE Trans. Pattern Anal. Mach. Intell.* 1–1 (2020) doi:10.1109/TPAMI.2020.3005590.

45. Liu, C. *et al.* Graph Pooling for Graph Neural Networks: Progress, Challenges, and Opportunities. Preprint at https://doi.org/10.48550/arXiv.2204.07321 (2023).

46. Funke, T., Khosla, M., Rathee, M. & Anand, A. ZORRO: Valid, Sparse, and Stable Explanations in Graph Neural Networks. *IEEE Trans. Knowl. Data Eng.* 1–12 (2023) doi:10.1109/TKDE.2022.3201170.

47. Hooker, S., Erhan, D., Kindermans, P.-J. & Kim, B. A Benchmark for Interpretability Methods in Deep Neural Networks. in *Advances in Neural Information Processing Systems* (eds Wallach, H. et al.) vol. 32 (Curran Associates, Inc., 2019).

48. Petersen, R. C. *et al.* Alzheimer's Disease NeuRoImaging Initiative (ADNI): Clinical characterization. *Neurology* **74**, 201–209 (2010).

49. Ellis, K. A. *et al.* The Australian Imaging, Biomarkers and Lifestyle (AIBL) study of aging: methodology and baseline characteristics of 1112 individuals recruited for a longitudinal study of Alzheimer's disease. *Int. Psychogeriatr.* **21**, 672–687 (2009).

50. LaMontagne, P. J. *et al.* OASIS-3: Longitudinal NeuRoImaging, Clinical, and Cognitive Dataset for Normal Aging and Alzheimer Disease. Preprint at https://doi.org/10.1101/2019.12.13.19014902 (2019).

51. Jack, C. R. *et al.* The Alzheimer's disease neuRoImaging initiative (ADNI): MRI methods. *J. Magn. Reson. Imaging* **27**, 685–691 (2008).

52. Zhou, H., He, L., Chen, B. Y., Shen, L. & Zhang, Y. Multi-Modal Diagnosis of Alzheimer's Disease Using Interpretable Graph Convolutional Networks. *IEEE Trans. Med. Imaging* **44**, 142–153 (2025).

53. Tascedda, S. *et al.* Advanced AI techniques for classifying Alzheimer's disease and mild cognitive impairment. *Front. Aging Neurosci.* **16**, 1488050 (2024).

54. Wang, Z. *et al.* Dynamic Multi-Task Graph Isomorphism Network for Classification of Alzheimer's Disease. *Appl. Sci.* **13**, 8433 (2023).

55. Garnier-Crussard, A., Cotton, F., Krolak-Salmon, P. & Chételat, G. White matter hyperintensities in Alzheimer's disease: Beyond vascular contribution. *Alzheimers Dement.* **19**, 3738–3748 (2023).

56. Bos, D. *et al.* Cerebral small vessel disease and the risk of dementia: A systematic review and meta-analysis of population-based evidence. *Alzheimers Dement.* **14**, 1482–1492 (2018).

57. Van Oostveen, W. M. & De Lange, E. C. M. Imaging Techniques in Alzheimer's Disease: A Review of Applications in Early Diagnosis and Longitudinal Monitoring. *Int. J. Mol. Sci.* **22**, 2110 (2021).